%% file: main.tex
\documentclass[runningheads]{llncs}
\usepackage[colorlinks=true,linkcolor=blue,citecolor=blue,urlcolor=blue]{hyperref}
\usepackage[T1]{fontenc}
\usepackage{graphicx}
\usepackage[utf8]{vietnam}
\usepackage[vietnamese, english]{babel}
\usepackage{amsmath}
\usepackage{algorithm}
\usepackage{algpseudocode}
\usepackage{marvosym} 
\usepackage{subcaption}
\usepackage{orcidlink}
\usepackage[section]{placeins}
\makeatletter
\renewcommand\subsection{\@startsection{subsection}{2}{\z@}%
  {0.25cm}
  {0.3cm}
  {\normalfont\normalsize\bfseries}}
\makeatother

\renewcommand{\figurename}{\textbf{Fig}}
\makeatletter
\renewcommand{\fnum@figure}[1]{\textbf{\figurename~\thefigure.}}
\makeatother
\usepackage{enumitem}
\setlist[itemize]{topsep=2pt,itemsep=1pt,parsep=0pt,partopsep=0pt}

\begin{document}
\title{Integrated Semantic and Temporal Alignment for Interactive Video Retrieval}
%
%
\author{
Thanh-Danh Luu\inst{1,3}\textsuperscript{*}\orcidlink{0009-0003-5632-6317} \and
Le-Vu Nguyen-Dinh\inst{1,3}\textsuperscript{*}\orcidlink{0009-0009-0405-7534} \and
Duy-Bao Bui\inst{1,3}\textsuperscript{*}\orcidlink{0009-0002-8257-8959} \and
Duc-Thien Tran\inst{1,3}\textsuperscript{*}\orcidlink{0009-0007-6463-5987} \and
Nam-Tien Le\inst{2,3}\textsuperscript{*}\orcidlink{0009-0002-7633-646X} \and
Tinh-Anh Nguyen-Nhu\inst{2,3}\textsuperscript{\Letter}\orcidlink{0009-0003-6414-8397}
}

\authorrunning{Thanh-Danh Luu et al.}
%
\institute{
University of Science, VNU-HCM, Ho Chi Minh City, Vietnam\\
\email{\{24280055,24120016,23122021\}@student.hcmus.edu.vn}, \email{tdthien2426@apcs.fitus.edu.vn} \and
University of Technology, VNU-HCM, Ho Chi Minh City, Vietnam\\
\email{\{tien.leaiemperor,anh.nguyennhu2306\}@hcmut.edu.vn} \and
Viet Nam National University, Ho Chi Minh City, Vietnam
}
\maketitle  
%
\begin{abstract}
The growing volume of video data and the introduction of complex retrieval challenges, such as the Temporal Retrieval and Alignment of Key Events (TRAKE) task at the Ho Chi Minh City AI Challenge 2025, expose critical limitations in existing systems. Many methodologies lack scalable, holistic architectures and rely on "frozen" embedding models that fail on out-of-knowledge (OOK) or real-world queries. This paper introduces the comprehensive video retrieval framework developed by team AIO\_Owlgorithms to address these gaps. Our system features an architecture integrating TransNetV2 for scene segmentation, BEiT-3 for visual embeddings in Milvus, and Gemini OCR for metadata in Elasticsearch. We propose two components: (1) \textbf{QUEST} (Query Understanding and External Search for Out-of-Knowledge Tasks), a two-branch framework that leverages a Large Language Model (LLM) for query rewriting and an external image search pathway to resolve OOK queries; and (2) \textbf{DANTE} (Dynamic Alignment of Narrative Temporal Events), a dynamic programming algorithm that efficiently solves the temporally-incoherent TRAKE task. These contributions form a robust and intelligent system that significantly advances the state-of-the-art in handling complex, real-world video search queries.
\end{abstract}

\begingroup
\renewcommand\thefootnote{$*$}
\footnotetext{These authors contributed equally to this paper.}
\endgroup
\begingroup
\renewcommand\thefootnote{\Letter}
\footnotetext{Corresponding author: anh.nguyennhu2306@hcmut.edu.vn}
\endgroup

\let\clearpage\relax

\input{introduction_short_short}
\input{DataPreprocessing}
\input{OverallSystem}

\makeatletter
\def\subsubsection{\@startsection{subsubsection}{3}{\z@}%
  {1.0ex plus 0.2ex minus .2ex}
  {0.5ex plus .2ex}
  {\normalfont\normalsize\bfseries}}%
\makeatother

\input{SystemUsage}
\input{conclusion}
\begin{credits}
    \subsubsection{\ackname} This research is supported by AI VIET NAM (aivietnam.edu.vn)    
\end{credits}

\bibliographystyle{splncs04}
\bibliography{references}
\newpage

\end{document}

%% file: introduction_short_short.tex
\section{Introduction}
The rapid expansion of multimedia data and the emergence of complex evaluation settings, notably the Temporal Retrieval and Alignment of Key Events (TRAKE) task~\cite{AIChallenge2025,paper16}, expose key limitations in current retrieval systems. Many methodologies prioritize individual components over holistic, scalable architectures and rely on embedding models with "frozen" knowledge that fail on Out-of-Knowledge (OOK) queries. Furthermore, handling temporally ordered events, as required by TRAKE, remains computationally inefficient for most systems.

To address these gaps, we, team \textbf{AIO\_Owlgorithms}, propose a comprehensive retrieval framework. Our modular architecture integrates TransNetV2~\cite{TransNetV2} for segmentation, Gemini~\cite{gemini25} for OCR, BEiT-3~\cite{BEiT3} for embeddings, and Milvus~\cite{Milvus} for vector indexing. We introduce two novel components: (1) \textbf{QUEST}~\cite{enhanced2024}, an LLM-based framework to resolve OOK queries; and (2) \textbf{DANTE}, a dynamic programming algorithm that efficiently solves the temporally-incoherent TRAKE task. \textbf{In summary, our primary contributions are:}
\begin{itemize}
    \item An end-to-end system architecture designed for scalability, maintainability, and seamless upgrades.
    \item The integration of an LLM within our \textbf{QUEST} framework, which employs a dual-branch strategy to resolve OOK entities by semantically rewriting queries and visually grounding them with external search results.
    \item A novel algorithm, \textbf{DANTE}, designed for the TRAKE task, which improves temporal alignment accuracy.
\end{itemize}
\section{Related Work}
Current research faces two significant challenges relevant to our work: bridging the "knowledge gap" of static models and efficiently processing complex temporal queries. \textbf{First}, a severe drawback of many systems is their reliance on embedding models with static, "frozen" knowledge bases. These models lack the real-world knowledge and reasoning capabilities to handle complex or OOK queries. Even when RAG is used, its application is often narrowly confined to searching over text transcripts, failing to bridge this fundamental knowledge gap. Our \textbf{QUEST} framework addresses this by using an LLM to rewrite queries and fetch external visual knowledge. \textbf{Second}, complex temporal event retrieval, as required by the TRAKE task, imposes a significant computational burden. Standard methods that search for multiple moments~\cite{nguyennhu2025stervlmspatiotemporalenhancedreference,nguyennhu2025lightweightmomentretrievalglobal,tran2025efficientrobustmomentretrieval} suffer from high latency and cost as data scales, rendering them inefficient for practical use. Our \textbf{DANTE} algorithm is proposed as an optimized solution to this specific problem.

%% file: DataPreprocessing.tex
\section{Data Preprocessing}

\begin{figure}
    \centering
    \includegraphics[width=0.85\textwidth]{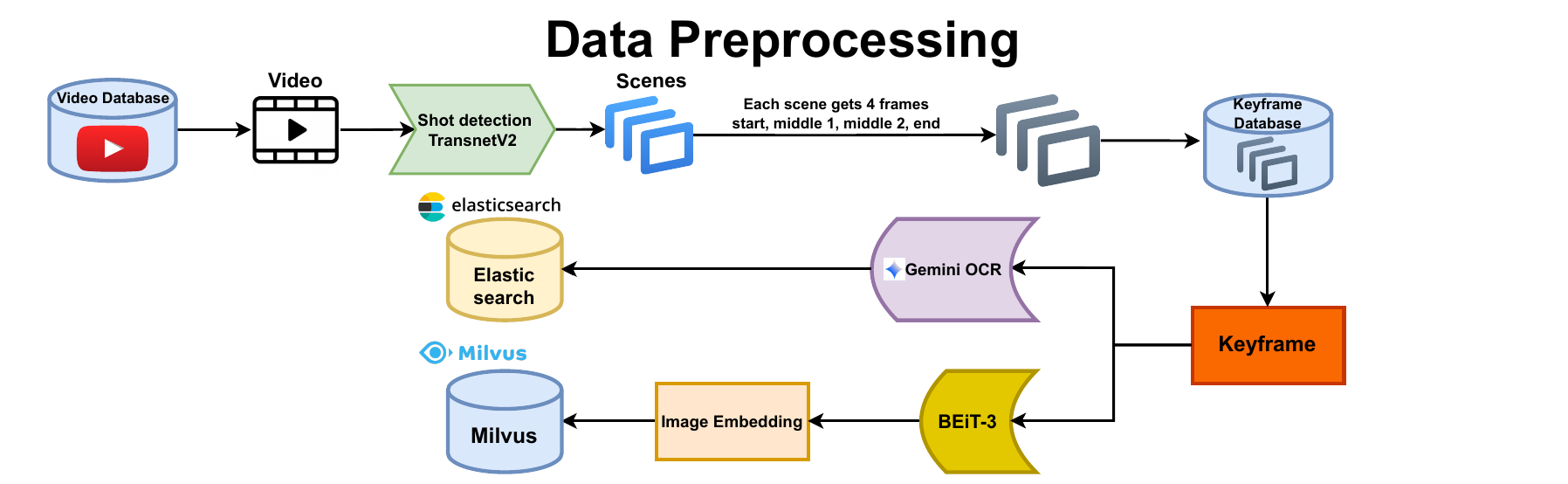}
    \caption{Overview of the data preprocessing pipeline, including shot detection, frame sampling, metadata extraction, and feature embedding.}
    \label{fig:Fig.1}
\end{figure}

All video data is converted into structured, searchable representations before retrieval. The preprocessing pipeline comprises three components—frame collection, metadata extraction, and visual feature embedding; see \textbf{Figure~\ref{fig:Fig.1}}.
\subsection{Frame Collection}
Videos are segmented into scenes using TransNetV2 to detect content transitions. Keyframes are extracted from each scene at four positions: the start, two from the middle, and the end, following the formula:
\[
k_{\text{extract}} = \{K_{a + \lfloor i \times (b - a)/3 \rfloor}, \forall i \in \{0, 1, 2, 3\}\}
\]
where \(K_j\) denotes the \(j\)-th frame in a scene spanning from frame \(a\) to \(b\). The keyframes are indexed with video identifiers and timestamps.
\subsection{Metadata Extraction}

Each keyframe is processed with Gemini to extract on-screen text (e.g., captions or labels), which is stored as key–value pairs (\texttt{keyframe\_id: ocr\_text}). This metadata is stored in an Elasticsearch database, with video-level details like timestamps and source identifiers linked to each keyframe for keyword-based filtering.
\subsection{Visual Feature Extraction}
Keyframes are converted into BEiT-3 visual embeddings that capture semantic and spatial information. To ensure efficiency, embedding is performed through a multi-stage, GPU-accelerated pipeline with dynamic batching. The resulting embeddings are precomputed, L2-normalized, and indexed in Milvus, enabling fast cosine-similarity retrieval without re-encoding at query time.

\subsection{Multimodal Indexing and Metadata Flow}
The system adopts a modular storage design anchored by a unified \texttt{keyframe\_id}. After extracting all keyframes from every scene, each keyframe is assigned a \texttt{keyframe\_id}, which serves as the canonical join key across all backends.\\

\noindent\textit{Storage Backends}
\begin{itemize}
\item \texttt{Vector (Milvus)}: \{keyframe\_id, embedding\_vector\}.
\item \texttt{OCR (Elasticsearch)}: \{keyframe\_id, ocr\_text\}.
\item \texttt{Metadata (MongoDB)}: \{keyframe\_id, image\_path\}; video-level \{video\_id, start index ($s_v$), end index($e_v$)\}.
\end{itemize}

\paragraph{Query-Time Hydration}
After retrieval modes (Semantic, OCR, and DANTE) output ranked lists of \texttt{keyframe\_ids}, which are resolved at query time by fetching their metadata (paths, timestamps, video info). This step standardizes result presentation and enables follow-up operations such as image-to-image search.

%% file: OverallSystem.tex
\section{Overall system}
This section presents an overview of the retrieval system, outlining its architecture and the integration of its core components, including embedding-based search, OCR search, and the two primary methodologies -- \textbf{DANTE} (Dynamic Alignment of Narrative Temporal Events)  and \textbf{QUEST} (Query Understanding and External Search for out-of-knowledge Tasks).
\begin{figure}[htbp]
    \centering
    \includegraphics[width=0.79\textwidth]{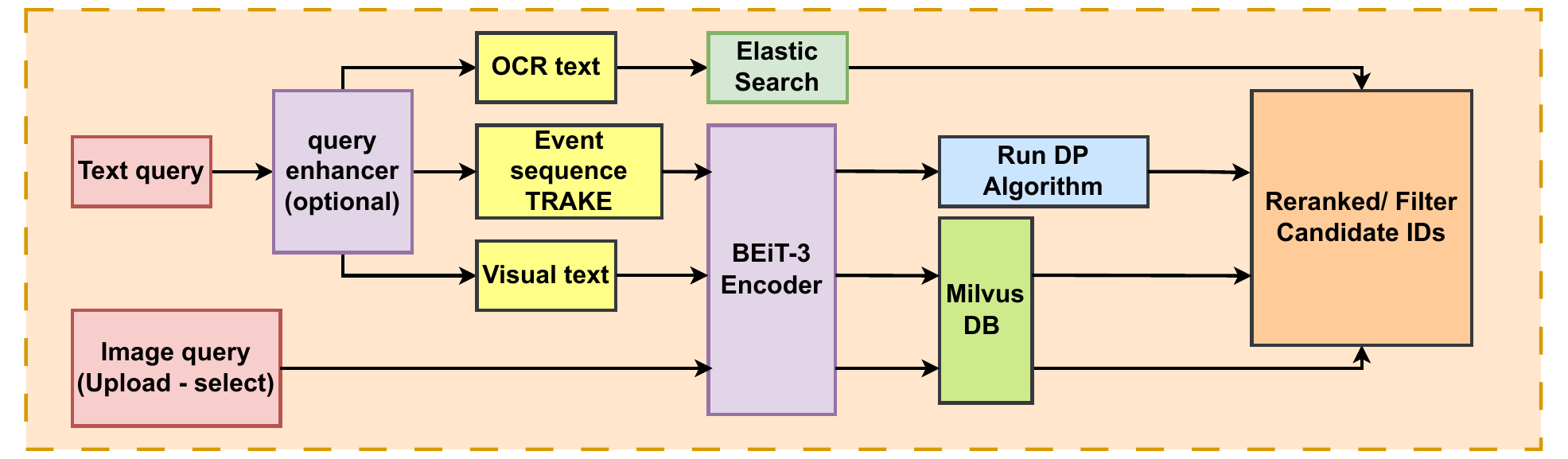}
    \caption{Overview of the online retrieval pipeline, where text or image queries are processed by optional LLM enhancers, the BEiT-3 encoder, or specialized modules (\textbf{DANTE}, OCR) before ranked results are returned from Milvus or Elasticsearch.}
    \label{fig:system-pipeline}
\end{figure}
\input{Overall_system_content/System}
\input{Overall_system_content/OOK}
\input{Overall_system_content/DANTE}

%% file: Overall_system_content/System.tex
\subsection{Base Search}
\label{sec:base_search}

The \textbf{Base Search} module supports two complementary retrieval modes: \textbf{Semantic Search}, which retrieves images by visual-semantic similarity, and \textbf{OCR Search}, which locates text appearing within images, together providing unified semantic and textual access to the dataset.

In \textbf{Semantic Search}, a user submits a text query describing the target visual content and specifies the Top-$k$ results to return. The query is encoded by BEiT-3, which maps textual
and visual representations into a unified multimodal embedding space, where the cosine similarity between the query embedding and precomputed image embeddings determines the most relevant results.

In \textbf{OCR Search}, a user inputs a text string to locate within images. The system uses the \emph{Vietnamese Analysis Plugin for Elasticsearch}~\cite{CocCocTokenizer,ElasticsearchVietnamese} to match the query against OCR text extracted by the Gemini API, enabling effective tokenization and similarity analysis for Vietnamese content. The Top-$k$ images with the highest similarity scores are then retrieved and ranked.

Additionally, users can refine retrieval by selecting an image from the initial results to perform an image-to-image search. The selected image’s serves as a new query, enabling retrieval of visually related images via cosine similarity and supporting the exploration of semantically similar content.

%% file: Overall_system_content/OOK.tex
\subsection{QUEST: Query Understanding and External Search for Out-of-Knowledge Tasks}\label{subsec:quest}

\begin{figure}[htbp]
    \centering
    \includegraphics[width=0.75\textwidth]{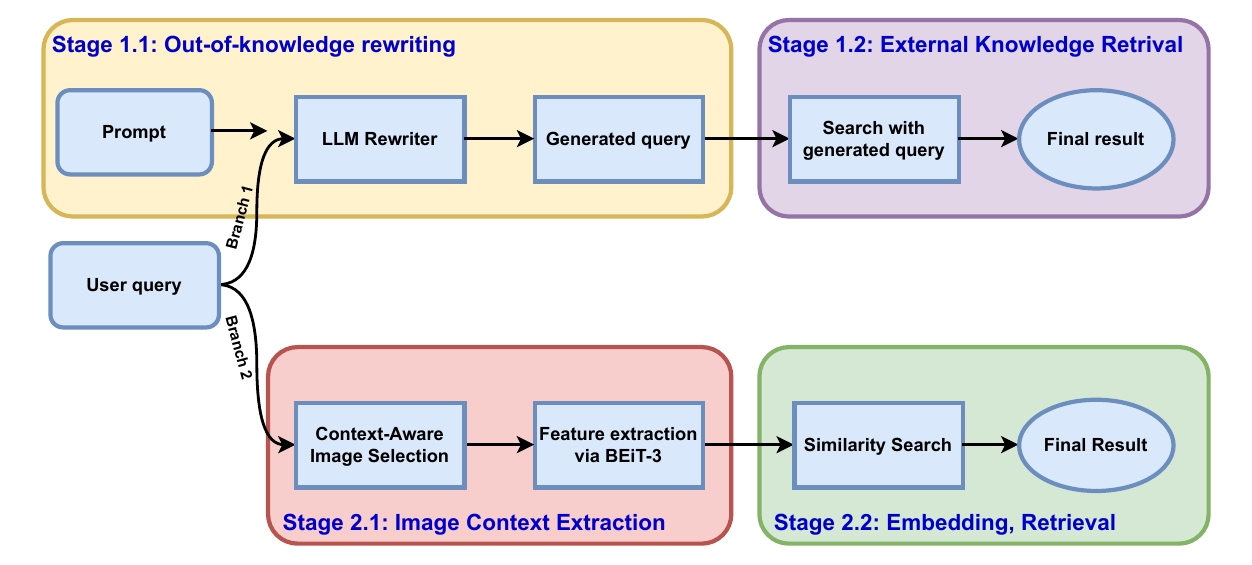}
    \caption{Overview of the \textbf{QUEST} framework.}
    \label{fig:quest_flow}
\end{figure}

\textbf{QUEST} is a two-branch framework designed to address unfamiliar or newly emerged visual entities by enhancing query interpretability through both language-based reasoning and external visual grounding; see \textbf{Figure}~\ref{fig:quest_flow}. From a single query, the system activates one of two complementary pathways depending on the nature of the input.

\noindent\textbf{Branch 1: Query Rewriting and System Retrieval.}
\textbf{QUEST} uses a large language model (e.g., GPT-4, Gemini) to rewrite the input query $q_{0}$ into a more descriptive, visually grounded form $q_{r}$. This rewritten query aligns better with the BEiT-3 embedding space and is processed through the standard retrieval pipeline (see \textbf{Section}~\ref{sec:base_search}), enabling improved performance for ambiguous or OOK textual descriptions.

\noindent\textbf{Branch 2: External Image Retrieval and Visual Matching.}
In parallel, \textbf{QUEST} may perform external image search following the open-image retrieval strategy in~\cite{enhanced2024}. A representative exemplar $I^{*}$ is selected (e.g., from Google Images) to visually anchor the concept. The exemplar is then encoded by BEiT-3 and used for image-to-image retrieval, which is particularly effective for novel or visually specific entities.

%% file: Overall_system_content/DANTE.tex
\subsection{\textbf{DANTE}: Dynamic Alignment of Narrative Temporal Events}
\label{sec:dante}
We introduce the \textbf{DANTE} (Dynamic Alignment of Narrative Temporal Events) algorithm, a novel framework that efficiently retrieves and aligns sequences of temporal moments in narrative order to handle TRAKE queries. This algorithm employs dynamic programming (DP) to optimize the search and alignment of consecutive keyframes in videos, leveraging the continuous indexing structure of keyframes in each video from the Milvus database.

\begin{figure}[htbp]
    \centering
    \includegraphics[width=0.7\linewidth]{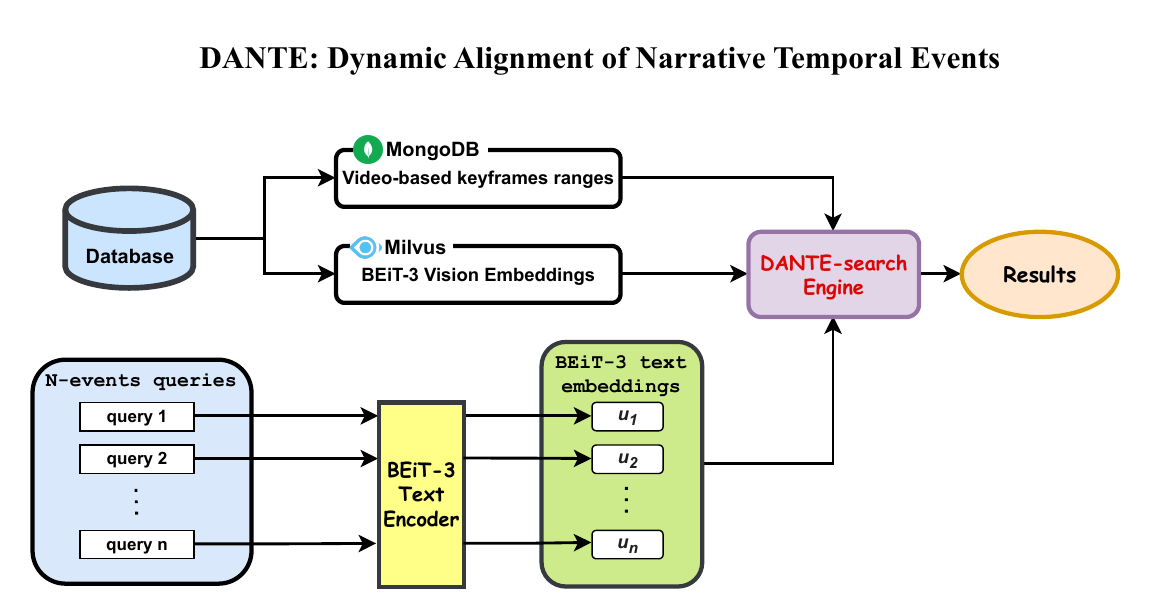}
    \caption{Illustration of \textbf{DANTE} workflow from input to output.}
    \label{fig:dante_pipeline}
\end{figure}

The algorithm takes as input the embedding vectors of \(N\) event queries (\(u_1, u_2, \ldots, u_N\)) embedded by BEiT-3, along with a JSON metadata file containing start index \(s_v\) and end index \(e_v\) for each video \(v\) (aggregated from \(V\) videos, such that the interval \([1, T]\) with \(T\) keyframes is covered by \(\cup [s_v, e_v]\)). The similarity matrix \(S[i, t]\) is precomputed using cosine similarity between \(u_i\) and the embedding of keyframe \(t\) from Milvus. The parameter \(\lambda\) is used to penalize temporal distances, helping prioritize nearby keyframes without completely eliminating farther options if they have high similarity. The pseudocode of the algorithm is presented specifically in Algorithm \ref{alg:dante}. \\
\begin{algorithm}[htbp]
\caption{\textbf{DANTE}: Dynamic Alignment of Narrative Temporal Events}
\label{alg:dante}
\begin{algorithmic}[1]
\Require Query embedding vectors \(U = [u_1, \ldots, u_N]\), JSON Metadata, Embeddings \(E[1..T]\), Penalty factor \(\lambda\)
\State Compute \(S[i, t] \gets \) Eq.\eqref{eq:0} // From Milvus
\For{each video \(v\) in Metadata}
    \State \(s_v, e_v = \text{Metadata}[v].\text{start}, \text{Metadata}[v].\text{end}\)
    \State Initialize \(DP[1..N, s_v..e_v]\)
    \For{\(t = s_v\) to \(e_v\)}  // Base case for \(i=1\)
        \State \(DP[1, t] = S[1, t]\)
    \EndFor
    \For{\(i = 2\) to \(N\)}
        \State \(running\_max \gets -\infty\)
        \For{\(t = s_v\) to \(e_v\)}
            \If{\(t > s_v\)}
                \State \(running\_max \gets \)Eq.\eqref{eq:2}

            \EndIf
            \State \(DP[i, t] \gets \)  Eq.\eqref{eq:3}
        \EndFor
    \EndFor
    \State \(DANTE[v] \gets\)Eq.\eqref{eq:4}
    \State Backtrack to retrieve keyframe sequence if needed 
\EndFor
\State Return top-\(k\) videos by \(DANTE[v]\), with backtracked keyframes
\end{algorithmic}
\end{algorithm}

\noindent The algorithm operates as follows: \\

\noindent \textbf{Step 1}: Compute the matrix 
\begin{equation}\label{eq:0}
S[i, t] = \text{cosine\_similarity}(u_i, E[t])
\end{equation}
for \(i = 1\) to \(N\) and \(t = 1\) to \(T\) (performed via Milvus queries to reduce load).  \\

\noindent\textbf{Step 2}: For each video \(v\), initialize the DP table \(DP[i, t]\) for \(t\in [s_v, e_v]\), representing the optimal score when matching event \(i\) to keyframe \(t\), considering previous matches and penalizing temporal distance:
  \begin{equation}\label{eq:1}
        DP[i, t] = S[i, t] + \max_{\tau \in [s_v, t-1]} \left( DP[i-1, \tau] - \lambda (t - \tau) \right)
  \end{equation}

 \noindent In this formula, the penalty factor $\lambda$ regulates the balance between temporal alignment and similarity strength, ensuring narrative order by preferring \(\tau\) close to $t$, but remains flexible with small \(\lambda\). During the final round of the Ho Chi Minh City AI Challenge 2025, we tuned~$\lambda$ in the range~0.001 to 0.01, achieving optimal performance across scenarios: $\lambda=0.001$ worked best when ground-truth query vectors~$(u_1, u_2, \ldots, u_N)$ had index gaps of~3–15, whereas~$\lambda=0.01$ was most effective for tighter alignments with~1–3 index offsets.\\

\noindent To optimize computation, we use a recursive form with running max:
  Initialize \(running\_max = -\infty\) (max of \(DP[i-1, \tau] + \lambda \tau\) for \(\tau \in [s_v, t-1]\)).
  Then, for each \(t \in ( s_v, e_v]\):
  \begin{equation}\label{eq:2}
 \quad running\_max = \max(running\_max, DP[i-1, t-1] + \lambda (t-1))
  \end{equation}
  \begin{equation}\label{eq:3}
        DP[i, t] = S[i, t] + running\_max - \lambda t
  \end{equation}
  This allows computation based solely on immediate previous values, ensuring complexity \(O(N (e_v - s_v + 1))\) per video, and equivalently $O(NT)$ over all keyframes. \\
  \textbf{Step 3}: Compute the DANTE score for each video \(v\): 
    \begin{equation}\label{eq:4}
  DANTE[v] = \max_{t \in [s_v, e_v]} DP[N, t] 
     \end{equation}
  \textbf{Output}: Top-\(k\) videos with the highest \textbf{DANTE} scores, along with sequences of \(N\) keyframes (one per query) obtained by backtracking from \(DP[N, t_{\text{max}}]\). Backtracking can be performed by storing argmax for each DP cell to reconstruct the optimal path.

%% file: SystemUsage.tex
\section{System usage}
This section provides an overview of the retrieval system, including its user interface and core search techniques for handling large event datasets. It also presents illustrative examples and summarizes our team’s performance and results in the Final Round of the HCMC AI Challange 2025.
\input{System_usage_content/OverallUI}
\input{System_usage_content/Base-search}
\input{System_usage_content/Quesk_usage}

\input{System_usage_content/Dante_usage}

In the Final Round of the HCMC AI Challenge 2025, our system achieved \textit{Outstanding} performance on TRAKE, \textit{Excellent} results on Textual KIS, and \textit{Very Good} results on both Video KIS and QA. Overall, it received an \textit{Excellent} final rating, reflecting strong robustness across diverse query types.

%% file: System_usage_content/OverallUI.tex
\subsection{Overall UI}

\textbf{Section~A} is the query input area (\texttt{textarea}) with an AI enhancer (E) using GPT-4o-mini to refine text queries~\cite{enhanced2024}.
\textbf{Section~B} allows search mode selection (``Semantic'', ``OCR'', \textbf{DANTE}) and parameter adjustment (Top-$K$, score threshold, or penalty weight).
\textbf{Section~C} displays results as relevance-sorted grids (Semantic/OCR/\textbf{DANTE}); clicking an image opens detailed metadata, OCR text, and action buttons (\textit{Play YouTube}, \textit{Find Similar Images}) for further searching.
\textbf{Section~D} provides metadata filters by video group or ID and supports image upload for local visual similarity search.
All text, image, and event embeddings are processed by the BEiT-3 model on the backend.

\begin{figure}[htbp]
    \centering
    \includegraphics[width=0.7\textwidth]{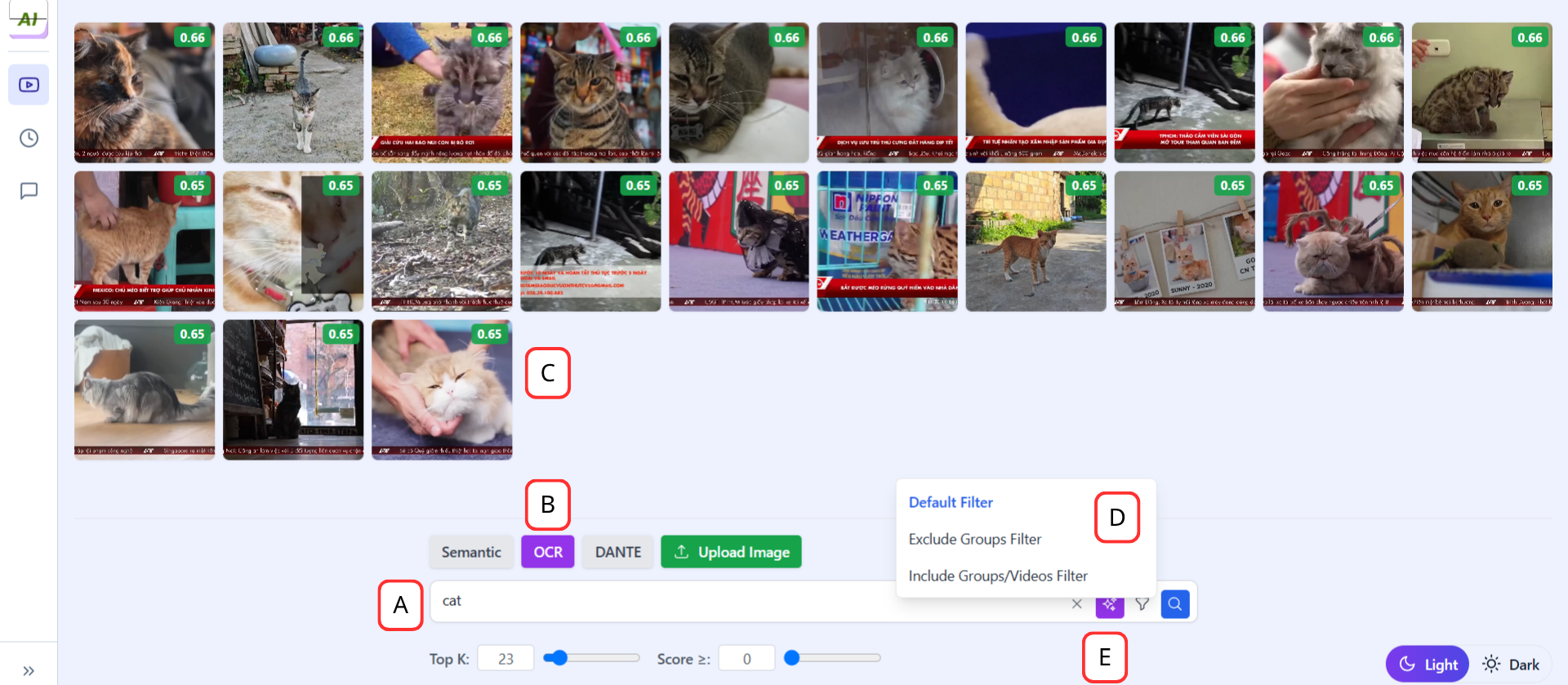}
    \caption{Overview of the multi-modal search UI, comprising query input (A) with optional AI enhancer (E), search type and parameter controls (B), results display (C), and pre-search metadata filters (D); supports image upload or 'Find Similar' functionality within (C).}
    \label{fig:}
\end{figure}

%% file: System_usage_content/Base-search.tex
\subsection{Semantic Search and OCR Search}
\label{subsec:base_usage}

This section illustrates how \textbf{Semantic Search} and \textbf{OCR Search} operate in
the interface using two examples drawn from the Final Round textual KIS tasks.

\begin{figure}[h]
    \centering
    \begin{subfigure}[t]{0.45\textwidth}
        \centering
        \includegraphics[width=\textwidth]{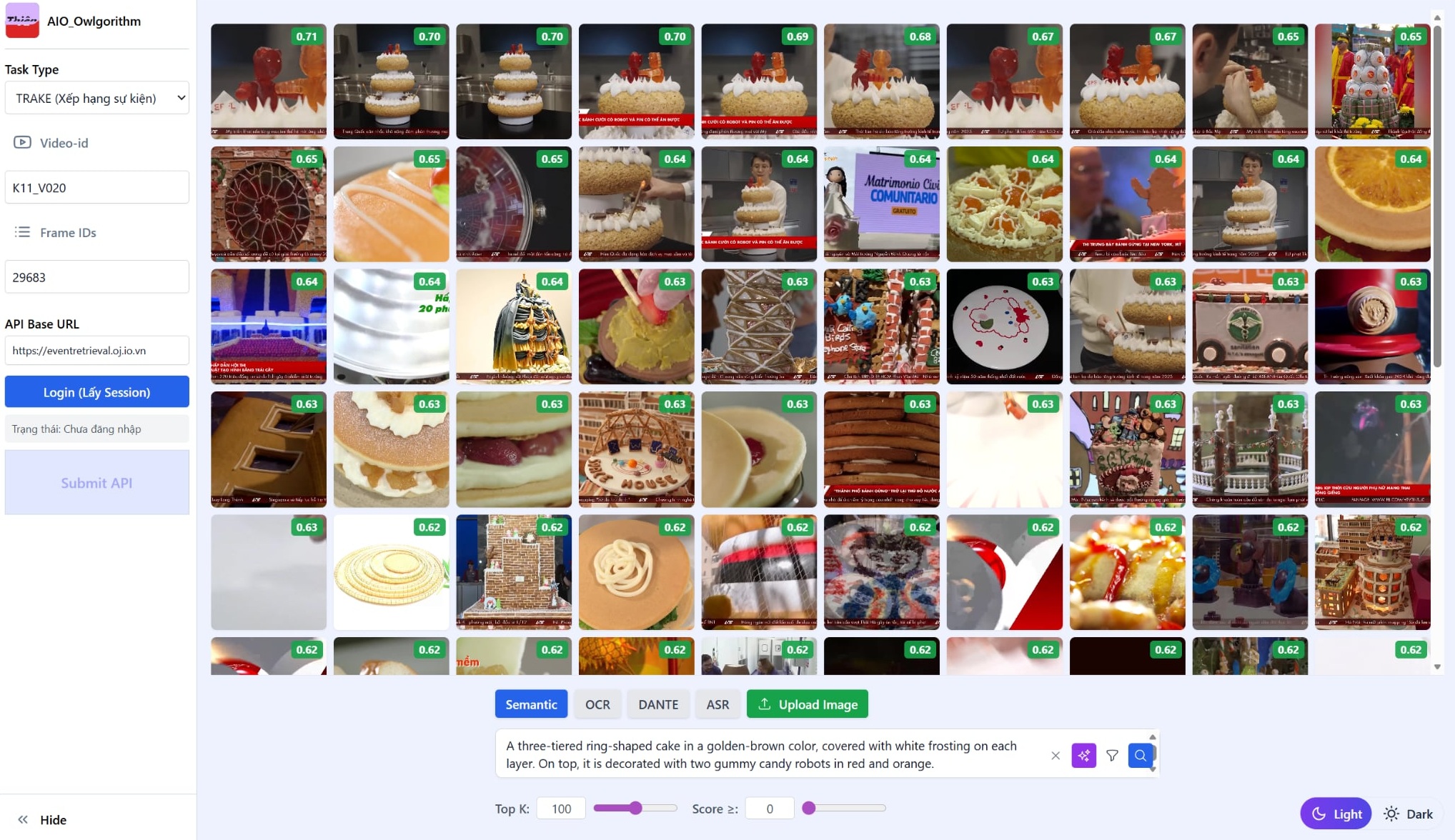}
        \caption{Semantic Search}
        \label{fig:semantic}
    \end{subfigure}
    \hspace*{\fill}%
    \begin{subfigure}[t]{0.45\textwidth}
        \centering
        \includegraphics[width=\textwidth]{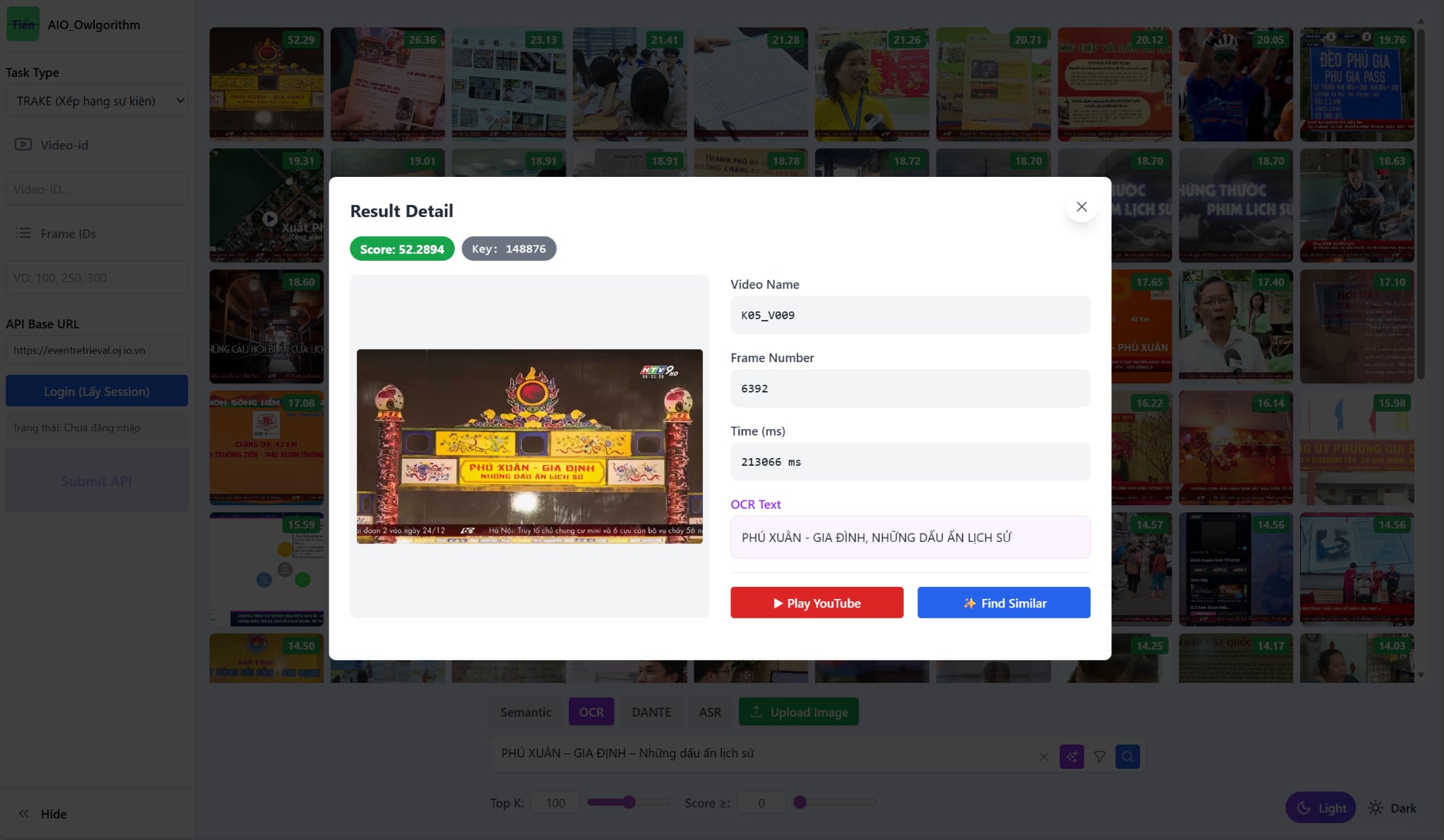}
        \caption{OCR Search}
        \label{fig:ocr}
    \end{subfigure}
    \caption{(a) Semantic Search and (b) OCR Search, demonstrating two complementary
    retrieval modes.}
    \label{fig:semantic-ocr}
\end{figure}

In the \textbf{Semantic Search} example (\textbf{Figure~\ref{fig:semantic}}), contestants entered a brief description of a three-tier cake with white frosting and two gummy robot toppers. The system retrieved the correct video frame at Top-1, showing its ability to interpret free-form descriptive queries. The \textbf{OCR Search} example (\textbf{Figure~\ref{fig:ocr}}) targeted a specific onscreen
phrase: ``PHÚ XUÂN – GIA ĐỊNH – những dấu ấn lịch sử.''
By matching OCR text extracted from keyframes, the system located the correct
instance immediately, also at Top-1.

%% file: System_usage_content/Quesk_usage.tex
\subsection{QUEST Usage} \label{subsec:quest_usage}
This section illustrates how the \textbf{QUEST} framework operates within the retrieval interface through two representative use cases corresponding to the branches in \textbf{Section~\ref{subsec:quest}}. Each example compares retrieval performance before and after enabling \textbf{QUEST}.


\begin{figure}[htbp]
    \centering
    \begin{subfigure}[t]{0.48\textwidth}
        \centering
        \includegraphics[width=0.8\textwidth]{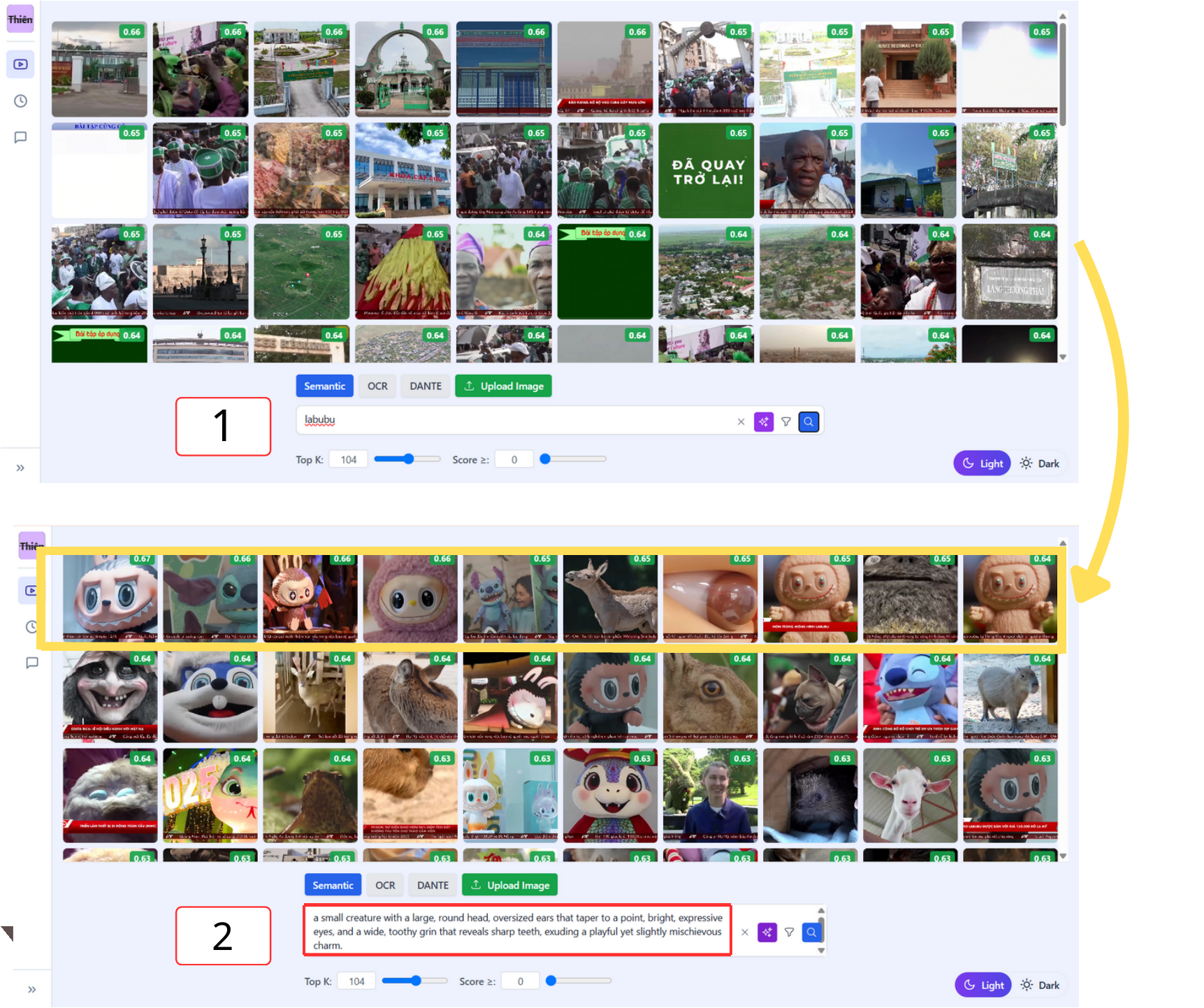}
        \caption{Query Rewriting mode in \textbf{QUEST}.  
        (1) Direct search for ``Labubu'' fails.
        (2) Rewriting expands the query and retrieves the correct Top-1 result.}
        \label{fig:quest_usage_branch1}
    \end{subfigure}
    \hspace*{\fill}%
    \begin{subfigure}[t]{0.435\textwidth}
        \centering
        \includegraphics[width=0.8\textwidth]{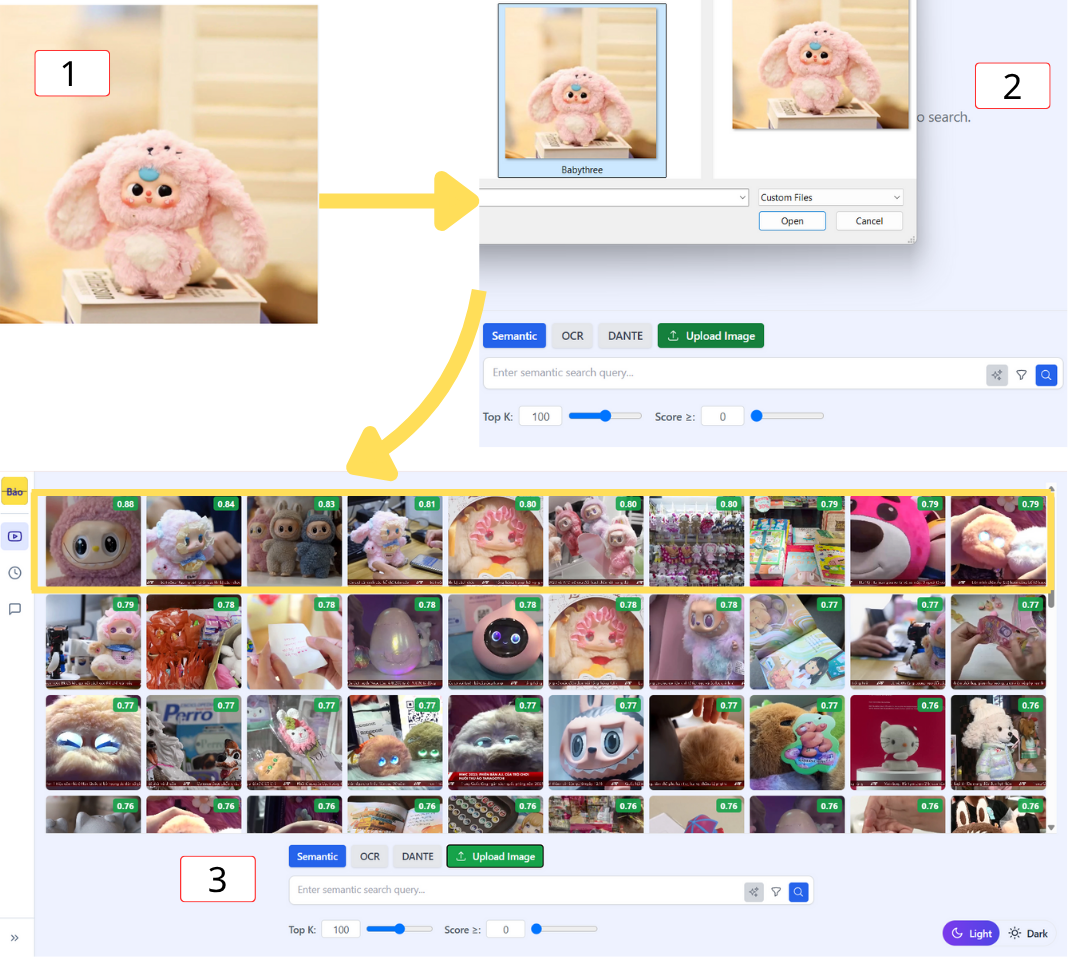}
        \caption{External Image Search mode.  
    (1) A representative web image of \textit{Babythree}.  
    (2) Uploaded as a visual query.  
    (3) Image-to-image retrieval returns the correct Top-1 result.}
        \label{fig:quest_usage_branch2}
    \end{subfigure}

    \caption{Illustration of the two operational modes of \textbf{QUEST}.  
    (a) Query Rewriting Mode. 
    (b) External Image Search Mode}
    \label{fig:quest-usage}
\end{figure}

\noindent In the \textbf{Query Rewriting Mode}, GPT-4o-mini reformulates the user query into a clearer, visually meaningful description, enabling retrieval of unseen concepts such as \textit{Labubu}; see \textbf{Figure~\ref{fig:quest_usage_branch1}}. \noindent In the \textbf{External Image Search Mode}, \textbf{QUEST} handles OOK queries by using a web image as visual input—e.g., an exemplar of \textit{Babythree} enables successful image-to-image retrieval where text alone fails (see \textbf{Figure~\ref{fig:quest_usage_branch2}}).

%% file: System_usage_content/Dante_usage.tex
\subsection{DANTE Usage Demonstration}

We illustrate the impact of \textbf{DANTE} across three representative scenarios:
one TRAKE-style multi-event query from the preliminary round and two Known-Item Search (KIS) tasks from the final round of the HCMC AI Challenge 2025.

\begin{figure}
    \centering
    \begin{subfigure}[t]{0.48\textwidth}
        \centering
        \includegraphics[width=\textwidth]{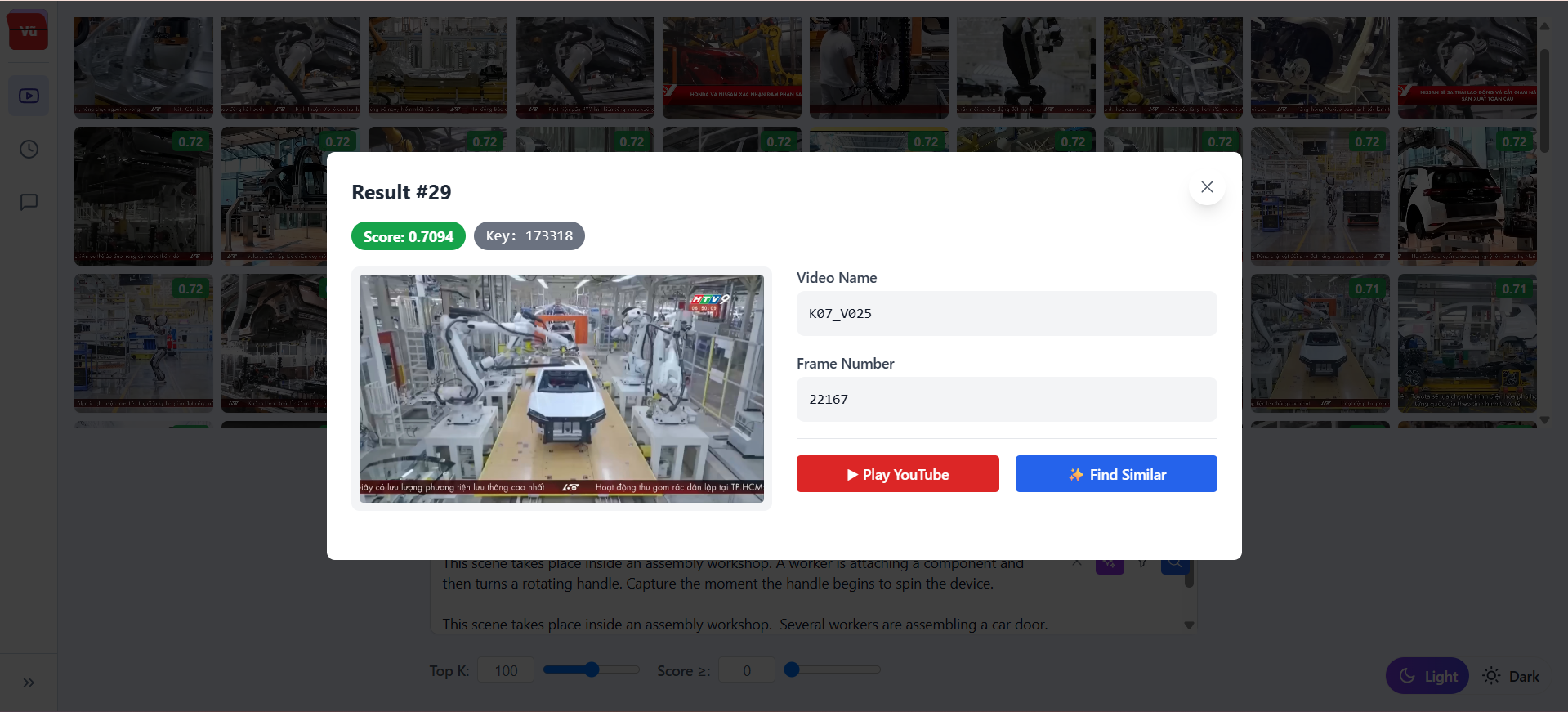}
        \caption{Semantic search.}
        \label{fig:trake-semantic}
    \end{subfigure}
    \hspace*{\fill}%
    \begin{subfigure}[t]{0.48\textwidth}
        \centering
        \includegraphics[width=\textwidth]{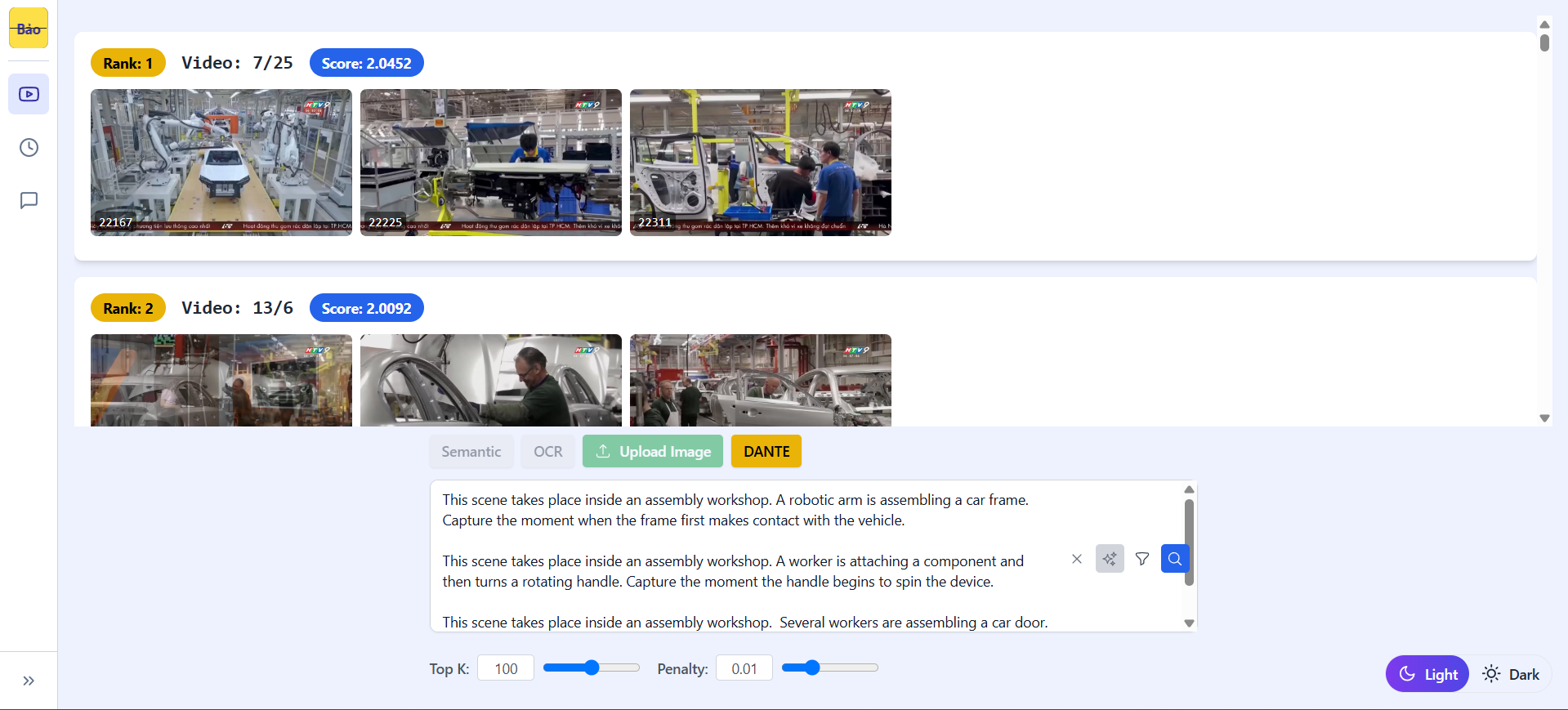}
        \caption{\textbf{DANTE}: aligned assembly sequence.}
        \label{fig:trake-dante}
    \end{subfigure}
    \caption{TRAKE-style assembly query showing the contrast between Semantic
    Search and \textbf{DANTE}.}
    \label{fig:trake-comp}
\end{figure}

As shown in \textbf{Figure~\ref{fig:trake-comp}}, the TRAKE-style query involves a set
of consecutive assembly-line actions. Semantic Search scatters relevant keyframes
(\textbf{Figure~\ref{fig:trake-semantic}}), placing the correct sequence deep in the
ranking, whereas \textbf{DANTE} reconstructs the full chronology
(\textbf{Figure~\ref{fig:trake-dante}}), demonstrating its ability to recover structured
multi-step activities.

\begin{figure}
    \centering

    \begin{subfigure}[t]{0.42\textwidth}
        \centering
        \includegraphics[width=\textwidth]{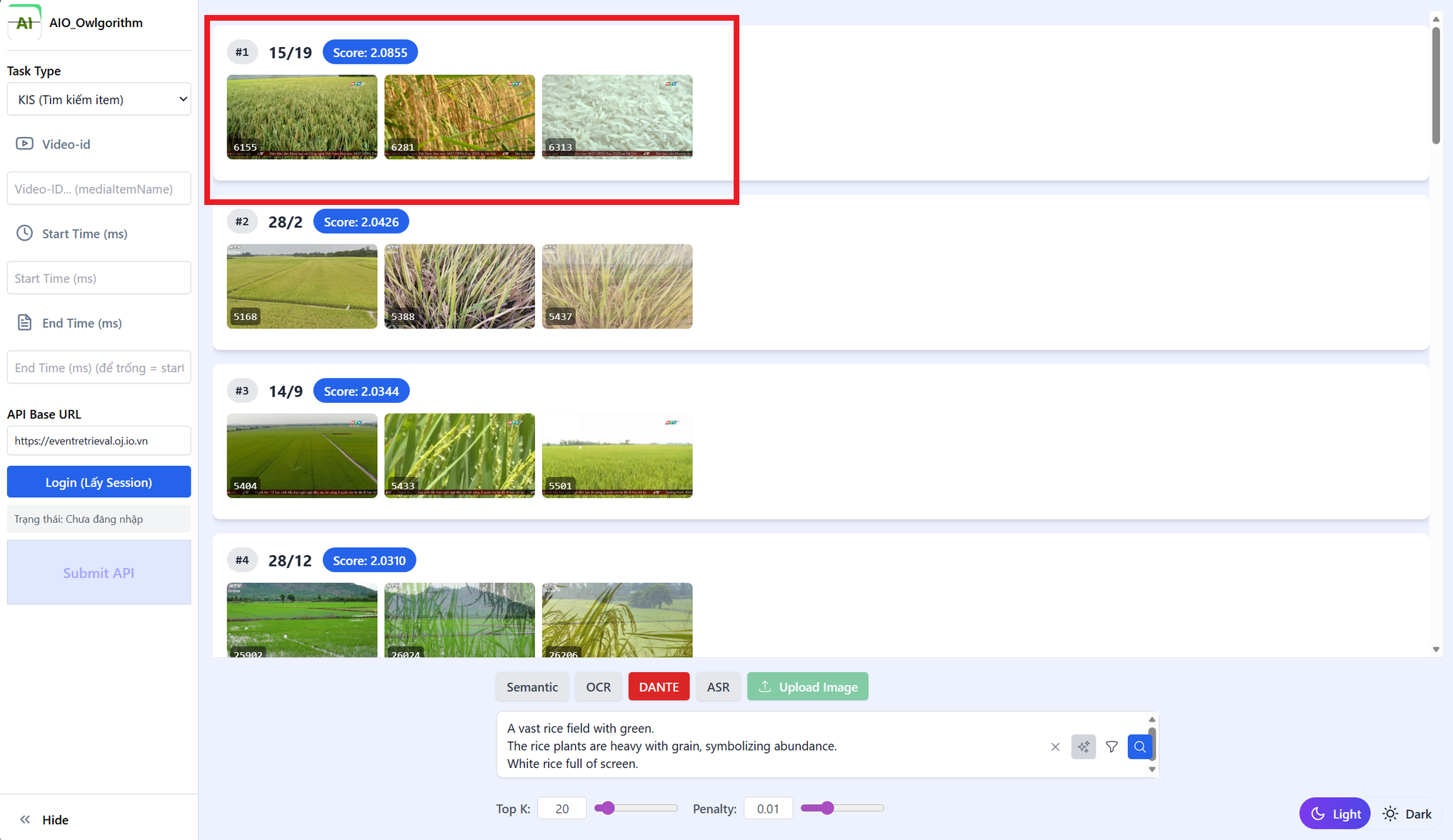}
        \caption{Video KIS: Top-1 result.}
        \label{fig:dante-vkis}
    \end{subfigure}
    \hspace*{\fill}%
    \begin{subfigure}[t]{0.42\textwidth}
        \centering
        \includegraphics[width=\textwidth]{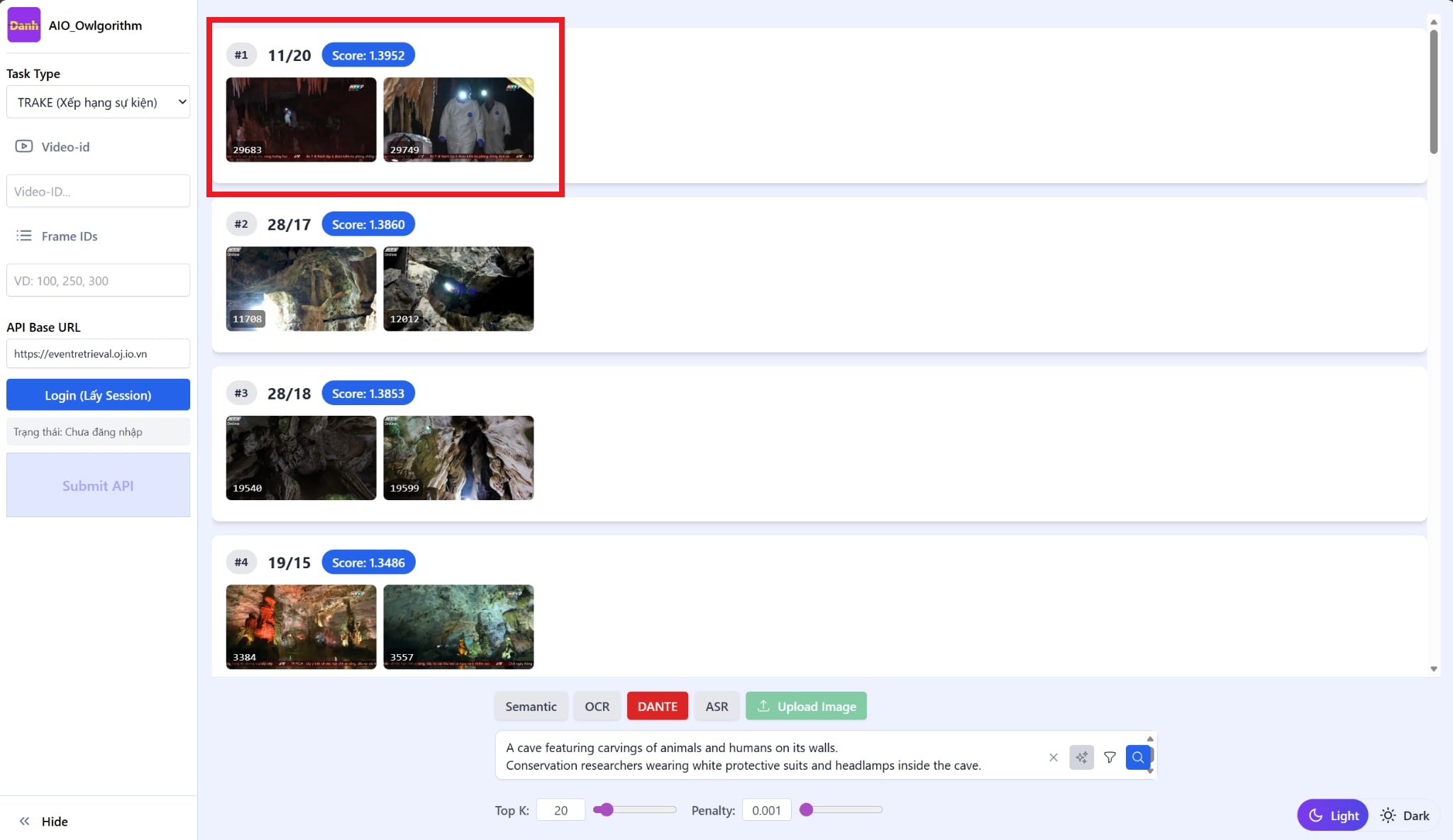}
        \caption{Textual KIS: Top-1 result.}
        \label{fig:dante-tkis}
    \end{subfigure}

    \caption{\textbf{DANTE} in two final-round KIS tasks. Ordered cues support
    consistent retrieval of the intended video.}
    \label{fig:dante-final}
\end{figure}

In the \textbf{Video KIS} task (\textbf{Figure~\ref{fig:dante-vkis}}), contestants viewed a
clip and provided three ordered descriptions:  
(1) a vast green rice field;  
(2) golden ripe rice bending;  
(3) close-up white rice grains.  
Semantic Search returned scattered frames, while \textbf{DANTE} aligned the events
and retrieved the correct video at Top-1.  In the \textbf{Textual KIS} task (\textbf{Figure~\ref{fig:dante-tkis}}), contestants typed two
descriptions of a news segment:  
(1) a cave in France with prehistoric animal and human carvings;  
(2) researchers in white protective suits with headlamps.  
Despite domain-specific phrasing, \textbf{DANTE} correctly combined both cues and
brought the target video to the leading position.

Overall, these examples show that \textbf{DANTE} improves retrieval not only in explicitly temporal tasks like TRAKE but also in KIS scenarios by enforcing coherent ordering, resulting in more reliable and user-aligned outputs.

%% file: conclusion.tex
\section{Conclusion}

In this paper, we, presented a holistic system developed for the HCMC AI Challenge 2025 to address key limitations in personal video retrieval. Our modular architecture integrates \textbf{QUEST} to overcome embedding knowledge gaps and \textbf{DANTE} to efficiently solve the fragmented TRAKE task, forming a robust and scalable framework for complex, real-world video search. Future work will explore lightweight adapter-based tuning or knowledge distillation to further improve efficiency. 

